\documentclass[a4paper,11pt]{article}
\usepackage{jinstpub} 
\usepackage{hyperref}
\usepackage{xcolor}

\title{\boldmath Evaluating the Effective Segregation Coefficient in High-Purity Germanium (HPGe) Crystals for Ge Detector Development in Rare-Event Searches}

\author{S. Chhetri $^a$, D.-M. Mei $^a$$^,$$^1$, S. Bhattarai $^a$, N. Budhathoki $^a$, A. Warren$^a$, K.-M. Dong$^a$, S.A. Panamaldeniya$^a$, A. Prem$^a$ }
\affiliation{$^a$Department of Physics, University of South Dakota, 414 E. Clark Street, Vermillion, South Dakota, 57069, USA\\}

\emailAdd{dongming.mei@usd.edu}

\abstract{
The performance and scalability of rare-event physics experiments depend on
large-volume, detector-grade high-purity germanium (HPGe) crystals with precise
control of impurity segregation during growth. We report a detailed study of
impurity distribution in a single Czochralski-grown HPGe crystal produced at
USD. The crystal was sectioned longitudinally into 37 segments, enabling the
first high-resolution and systematic mapping of dopant profiles along the length
of a detector-grade HPGe boule. Hall-effect measurements were used to extract
impurity concentrations for boron (B), aluminum (Al), gallium (Ga), and
phosphorus (P) in each segment. From these data, we determine effective
segregation coefficients ($K_\mathrm{eff}$) and initial melt concentrations
($C_0$) for the dominant dopants and compare them with classical
Burton–Prim–Slichter expectations. The results provide quantitative insight
into impurity transport and melt–solid partitioning under realistic detector
growth conditions. These findings inform process-optimization strategies for
HPGe crystal pulling, improve impurity control along the boule, and support the
reliable fabrication of large, low-background HPGe detectors for next-generation
rare-event searches.
}


\footnotetext[1]{Corresponding author.}

\begin{document}
\maketitle
\flushbottom

\section{Introduction}
\label{sec:intro}

Dark matter and neutrinos are among the most abundant yet least understood
constituents of the universe. High-purity germanium (HPGe) detectors, with net
impurity concentrations at the level of $\sim 10^{10}\,\mathrm{cm}^{-3}$, have
become key technologies for probing physics beyond the Standard Model, including   
dark-matter interactions and neutrinoless double-beta decay
\cite{anderson2011coherent,wei2017cosmogenic,agostini2015production,agostini2019probing,agnese2018nuclear,kermaidic201876ge,aalseth2013cogent,angloher2009commissioning}.  
Early germanium-based experiments such as CoGeNT \cite{aalseth2011search}, which
reported a possible dark-matter signal, and KKDC \cite{klapdor2002status}, which
claimed evidence for neutrinoless double-beta decay, helped establish Ge as a
leading target for rare-event searches. More sensitive experiments, including
SuperCDMS, EDELWEISS and CDEX
\cite{agnese2018first,armengaud2019searching,yang2019search}, have since ruled
out the CoGeNT-favored parameter space, yet Ge detectors still provide
state-of-the-art sensitivity for dark-matter masses below a few GeV/c$^{2}$. In
the double-beta decay sector, GERDA and MAJORANA
\cite{gerda2018upgrade,alvis2019search} have refuted the KKDC claim using large
arrays of ultra-low-background HPGe detectors.

Building on these advances, the LEGEND program \cite{romo2025results} aims at a
ton-scale search for neutrinoless double-beta decay using enriched HPGe. At the
same time, leading dark-matter experiments in the GeV mass range, such as
SuperCDMS \cite{agnese2018low}, LUX-ZEPLINE (LZ) \cite{aalbers2023first} and
XENONnT \cite{aprile2023first}, continue to push direct-detection limits.
Recently, the Germanium-based Quantum Sensors for Low-Energy Physics (GeQuLEP)
platform has been proposed, in which cryogenic (4 K) phonon spectroscopy with
HPGe enables sensitivity to MeV-scale dark matter \cite{mei2025probing,mei2025phonon}. In all
of these efforts, the sensitivity of Ge-based experiments depends critically on
the availability of large-volume, detector-grade HPGe crystals with well
controlled impurity content and uniformity.

Increasing the HPGe crystal diameter beyond 4 inches enhances rare-event
sensitivity, especially for weakly interacting massive particles (WIMPs).
However, crystals grown at the Earth's surface without adequate shielding are
exposed to cosmic rays that produce long-lived cosmogenic isotopes in the
crystal volume. These isotopes contribute to background events and degrade both
detector sensitivity and energy resolution. To mitigate such effects, the
entire production chain -- zone refining, crystal growth and detector
fabrication -- should ideally be carried out in an underground environment,
where the cosmic-ray flux is strongly suppressed \cite{mei2009cosmogenic}.
While underground crystal growth is technically and logistically challenging,
it is a key step toward achieving the ultra-low-background HPGe detectors
required for next-generation dark-matter and neutrinoless double-beta decay
experiments.

Detector-grade HPGe crystals are typically produced through a sequence of
highly controlled steps \cite{hansen1971high,haller1981physics,wang2015crystal}.
Raw Ge is first purified by zone refining \cite{yang2014investigation}, then
further purified during Czochralski (CZ) crystal pulling, and finally
characterized by Hall-effect measurements before detector fabrication
\cite{raut2020characterization}. A high-purity hydrogen atmosphere during zone
refining and crystal growth is essential \cite{hansen1971high,haller1981physics}
to suppress electrically active impurities. Detector-grade quality further
requires dislocation densities in the range
$3\times10^{2}$--$1\times10^{4}\,\mathrm{cm}^{-2}$, so that charge trapping by
di-vacancy–hydrogen complexes ($V_2$H) is minimized, and a net impurity
concentration of order $10^{10}\,\mathrm{cm}^{-3}$ with good homogeneity
throughout the volume \cite{bhattarai2024investigating,haller2007origin}.
Dislocation-free Ge crystals up to 30 cm in diameter can be grown in inert
atmospheres using resistance heating and graphite crucibles
\cite{depuydt2006germanium}. In contrast, CZ growth from ultra-pure quartz
crucibles in high-purity hydrogen with radio-frequency induction heating offers
improved thermal-field uniformity and strong melt convection, but also
introduces additional challenges: reactions between molten Ge and quartz can
create oxygen-related complexes that act as trapping centers, and ambient
hydrogen can form $V_2$H complexes that degrade charge collection
\cite{mei2020impact}.

Impurity segregation during zone refining and CZ growth is governed by the
equilibrium distribution coefficient
$K_0 = C/C_0$, defined as the ratio of the solute concentration in the solid
($C$) to that in the liquid ($C_0$) \cite{burton1953distribution,hall1953segregation}.
For HPGe, the dominant electrically active impurities are boron (B), aluminum
(Al), gallium (Ga) and phosphorus (P). Their segregation behavior controls how
these dopants are distributed along the length of the ingot or crystal: species
with $K > 1$ concentrate near the head, while those with $K < 1$ accumulate
near the tail \cite{wang2015crystal,yang2014investigation,bhattarai2024investigating}.
In practical growth conditions, however, convection, incomplete mixing and
finite growth rates lead to effective segregation coefficients
$K_\mathrm{eff}$ that can differ significantly from the equilibrium values.
Accurate knowledge of $K_\mathrm{eff}$ under realistic detector-growth
conditions is therefore essential for predicting impurity profiles and
optimizing HPGe crystal pulling.

\begin{figure}[h]
    \centering
    \includegraphics[width=1\textwidth]{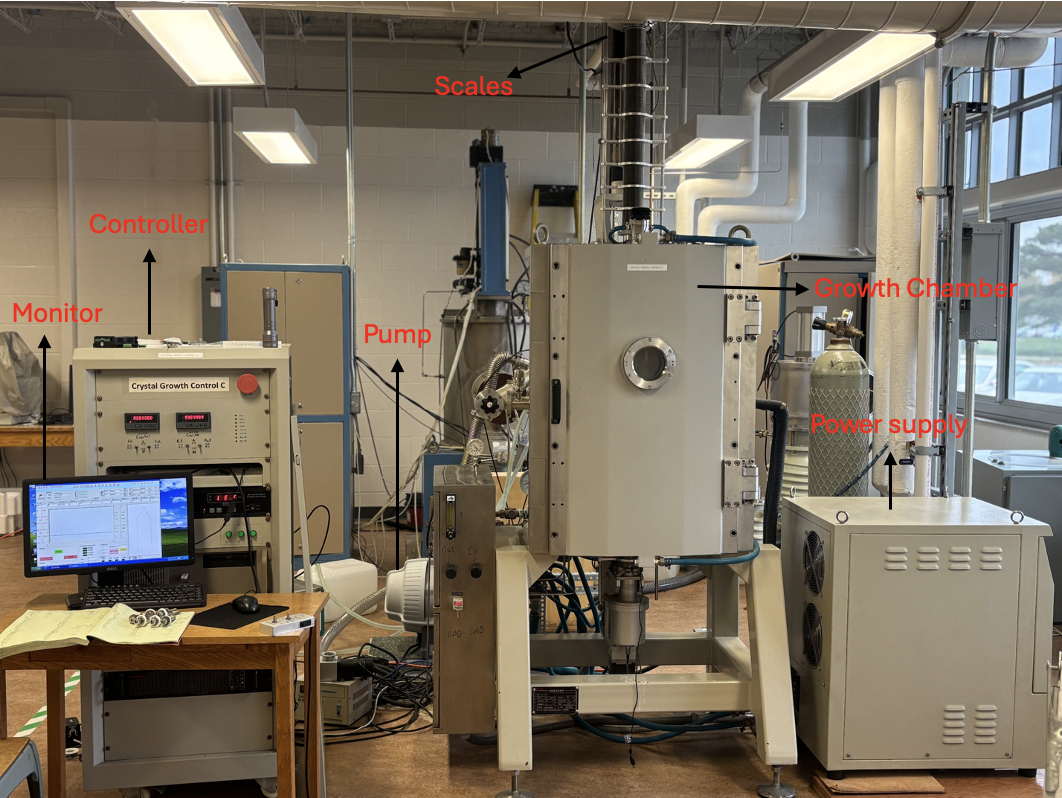}  
    \caption{Czochralski crystal growth chamber used at USD for high-purity
    germanium crystals under controlled thermal and atmospheric conditions.}
    \label{fig:1}
\end{figure}

The University of South Dakota (USD) has produced HPGe crystals using the
Czochralski technique for several decades and has developed the capability to
fabricate radiation detectors with various geometries from these crystals
\cite{wang2015high,bhattarai2020investigation,meng2019fabrication}. In the
present work, a single detector-grade HPGe crystal grown at USD was
longitudinally sectioned into 37 segments to obtain a high-resolution map of
impurity distribution along the boule. Using Hall-effect measurements, we
extract impurity concentrations for B, Al, Ga and P in each segment and
determine the corresponding effective segregation coefficients
$K_\mathrm{eff}$ and initial melt concentrations $C_0$. To our knowledge, this
is the first systematic, high-resolution longitudinal study of impurity
segregation in a single HPGe crystal grown under realistic detector conditions
at USD. The Czochralski growth chamber used in this work is shown in
Fig.~\ref{fig:1}; the methodology and results of the impurity-distribution
analysis are presented in the following sections.

\section{Experimental Procedures}
\label{sec:exp}

At USD, raw germanium with an initial impurity concentration of
$\sim 10^{14}~\mathrm{cm^{-3}}$ is first purified by a multi-pass zone-refining
process. After 14 passes, the net carrier concentration is reduced to
$\sim (10^{11}\text{--}10^{12})~\mathrm{cm^{-3}}$ as measured by Hall-effect
characterization at liquid-nitrogen temperature (77 K). The purified material
is then loaded into a high-purity quartz crucible, 18 cm in diameter and
11 cm in depth, for subsequent Czochralski (CZ) growth
\cite{yang2014investigation,yang2015zone}.

Before each growth run, a rigorous cleaning protocol is implemented to minimize
contamination. The Ge ingots are rinsed with deionized (DI) water and acetone
to remove organic residues, followed by a DI water rinse and a mixed-acid etch
in nitric and hydrofluoric acids in a (3:1) HNO$_3$:HF ratio to remove surface-bound
impurities. The quartz crucible and radiation shields are cleaned with
H$_2$O$_2$ and dried with high-purity nitrogen gas
\cite{bhattarai2024investigating}. The inner surfaces of the growth chamber
(Fig.~\ref{fig:1}) are also wiped and inspected to ensure that residues from
previous runs are removed, providing a clean environment for detector-grade
crystal growth.

A total of 4905 g of purified Ge is charged into the quartz crucible, which is
placed on an ultra-pure graphite susceptor that converts radio-frequency (RF)
power into heat. The charge is melted by applying $\sim 12.5$ kW of continuous
RF power for about 2 h. During this stage, the chamber is first evacuated with
a mechanical pump and then with a diffusion pump, improving the pressure from
$(6\text{--}7)$ Pa down to $(1\text{--}2)\times 10^{-3}$ Pa. After evacuation,
5N-purity hydrogen is introduced at a flow rate of 150 L/h and a pressure of
1 atm, and the atmosphere is maintained under these conditions throughout
growth.

\begin{figure}[h]
    \centering
    \includegraphics[width=1\textwidth]{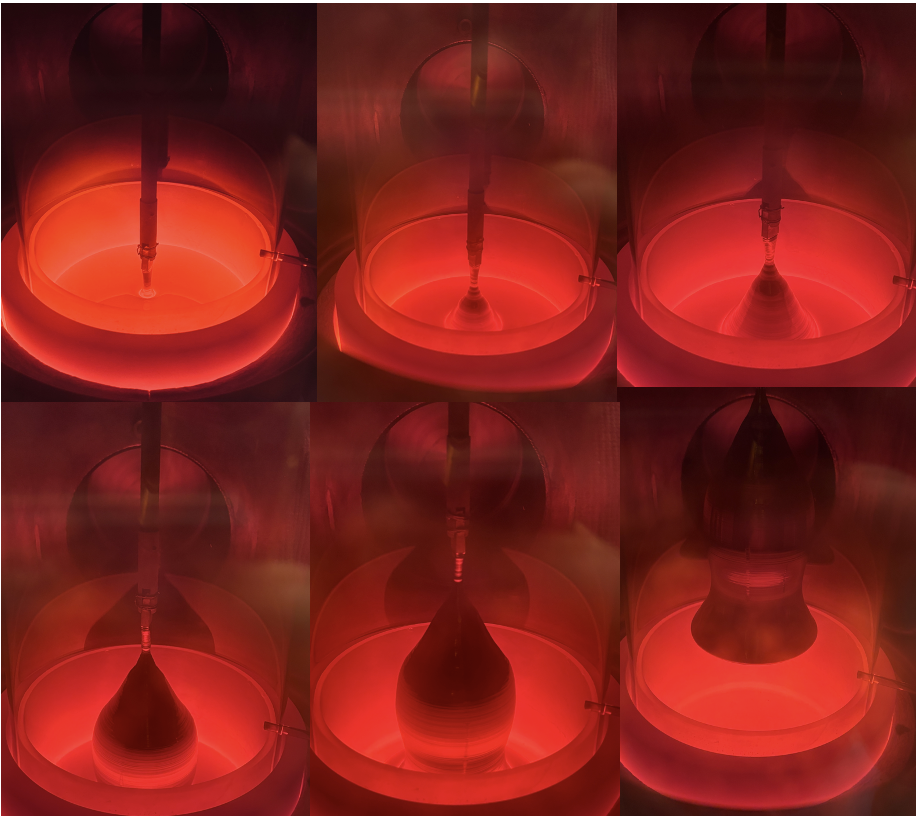}
    \caption{Schematic of the main stages of Czochralski growth for HPGe:
    (a) neck formation, (b) shoulder expansion, and (c) constant-diameter body
    growth.}
    \label{fig:2}
\end{figure}

Once the melt is fully established, a $\langle 100 \rangle$-oriented HPGe seed
is dipped into the melt and pulled to form the neck. The crystal is rotated at
3 rpm. The pulling rate is maintained at 40 mm/h during necking to suppress
dislocations and then reduced to 20 mm/h for shoulder enlargement and
constant-diameter body growth, as illustrated in Fig.~\ref{fig:2}. After
melting, the RF power is gradually reduced to about 9.5 kW and subsequently
adjusted to maintain the desired crystal diameter and a stable solid–liquid
interface. Particular care is taken in the final stage of growth to avoid
leaving residual molten Ge in the crucible, which can impose excessive thermal
stress and crack the quartz.

\begin{figure}[h]
    \centering
    \includegraphics[width=1\textwidth]{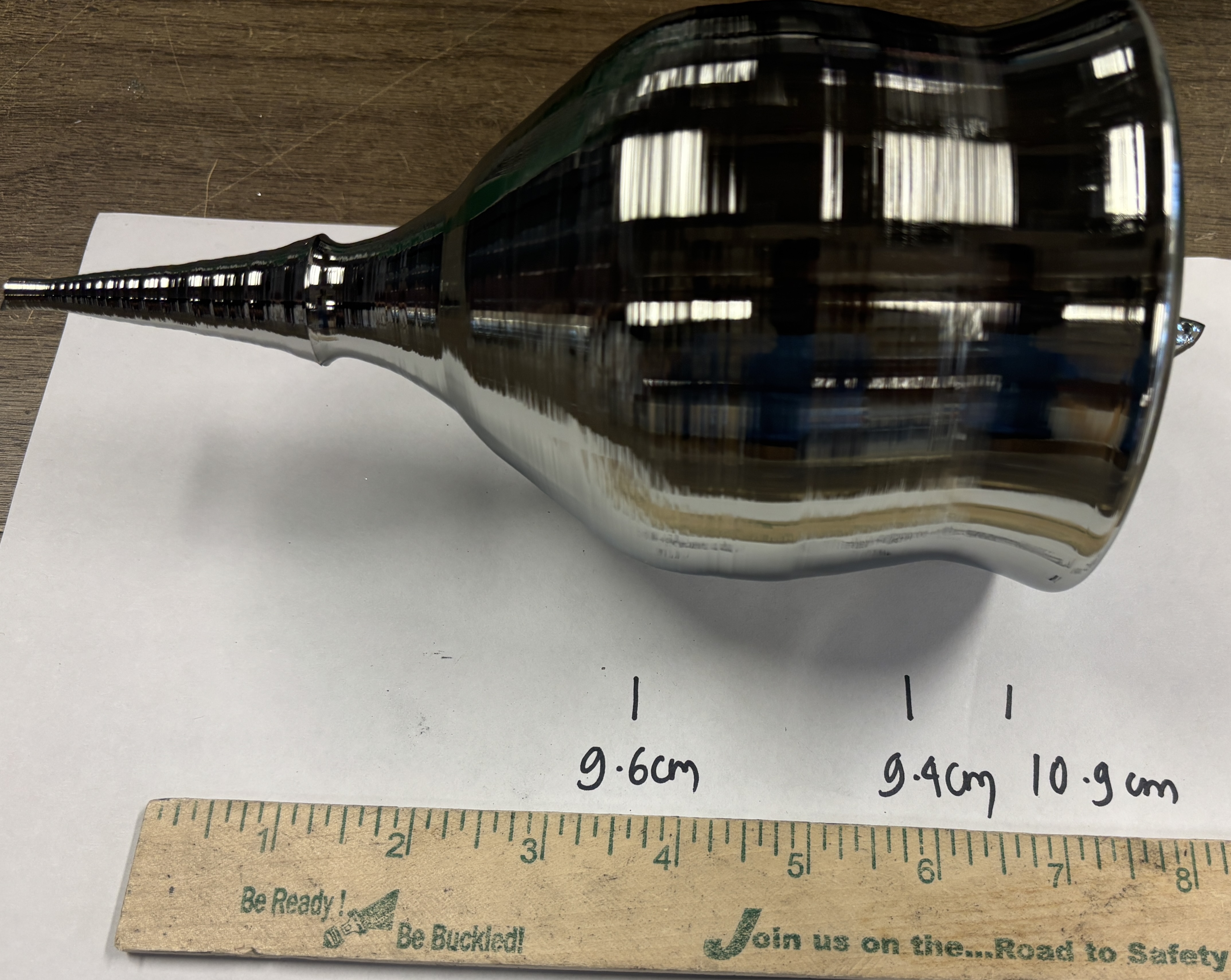}
    \caption{HPGe crystal grown at USD using the Czochralski technique. The
    diameter is $\sim 8$ cm near the shoulder and increases to $\sim 11$ cm
    toward the tail region.}
    \label{fig:3}
\end{figure}

Upon completion of the pull, a controlled automated cool-down of
$\sim 10$–11 h is used to bring the system back to room temperature and to
minimize thermal stress in both crystal and crucible. The resulting
$\langle 100 \rangle$ HPGe crystal, shown in Fig.~\ref{fig:3}, has a diameter
of approximately 8 cm near the shoulder, expanding to about 11 cm toward the
tail, and is then removed from the chamber for slicing and subsequent
characterization.

\section{Characterization}
\label{sec:characterization}

\begin{figure}[h]
\centering
    \includegraphics[width=1\textwidth]{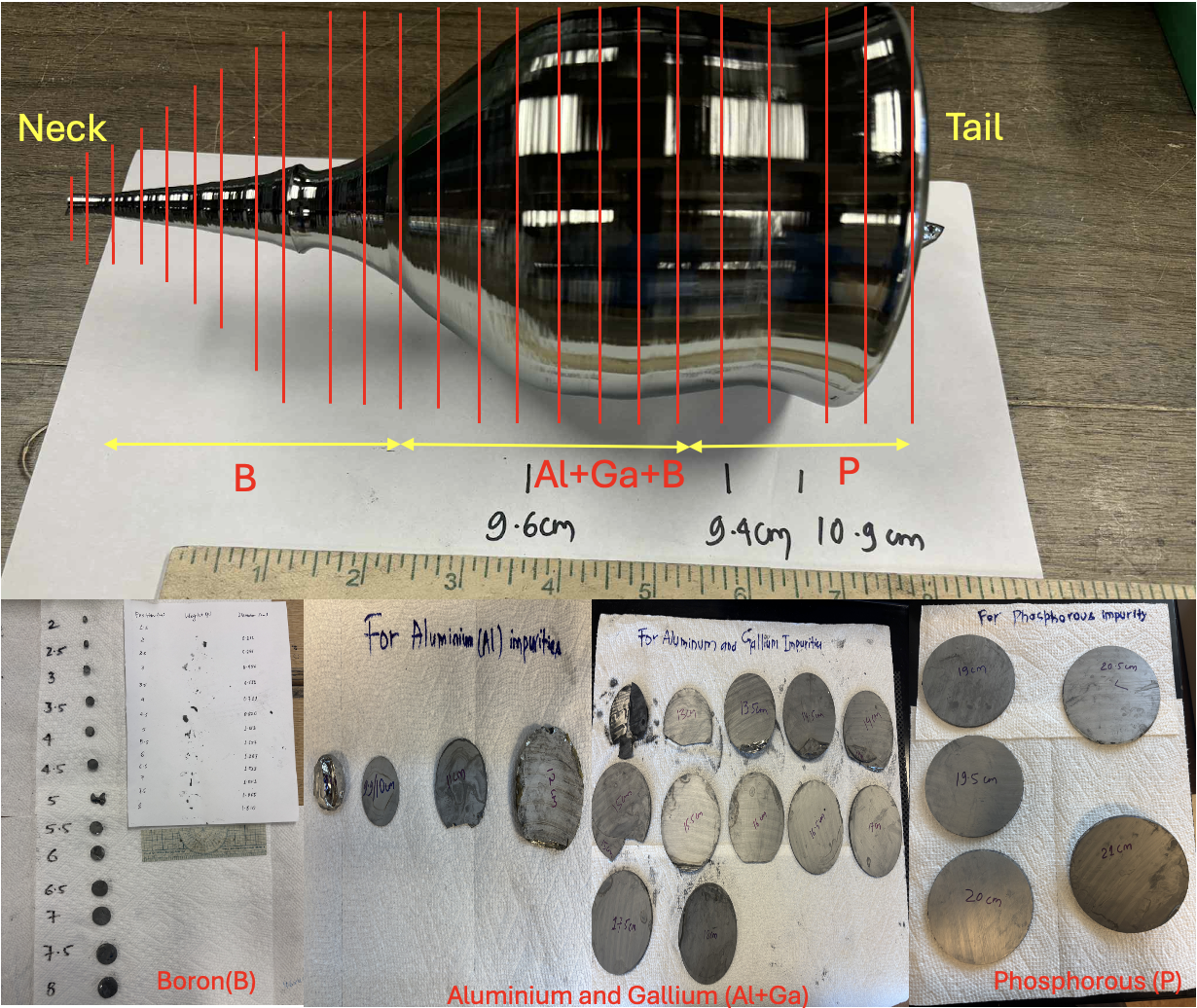}
    \caption{Impurity distribution along the length of the HPGe crystal grown at USD. Red vertical lines mark the positions where the crystal was longitudinally sectioned into 37 segments using a diamond wire saw for Hall-effect characterization. The profile shows strong boron dominance in the first $\sim 15$ cm of the crystal.}
    \label{fig:4}
\end{figure}

The single HPGe crystal was longitudinally sectioned into 37 segments to obtain
a high-resolution impurity profile along the boule, as shown in
Fig.~\ref{fig:4}. Sectioning was performed with a diamond wire saw, after which
each segment was trimmed to the required dimensions using a water-cooled saw
and finished on a lapping wheel to achieve flat, uniform surfaces suitable for
electrical characterization.

\begin{figure}[h]
\centering
    \includegraphics[width=1\textwidth]{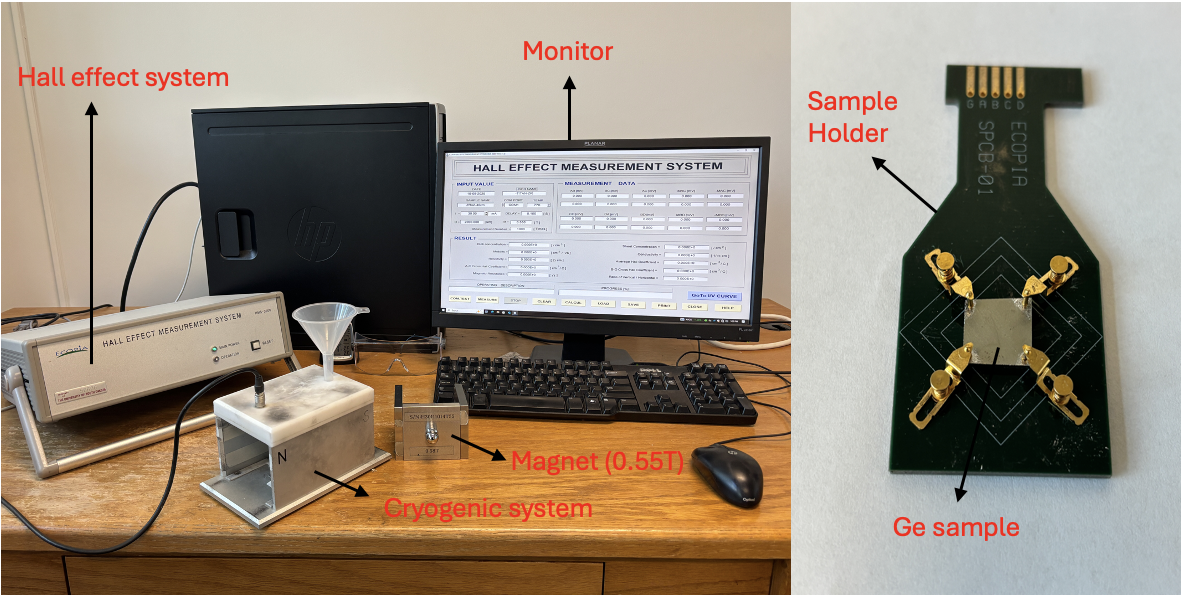}
    \caption{Ecopia HMS-3000 Hall-effect measurement system at USD used to
    determine electrical properties and net impurity concentrations of HPGe
    samples at 77 K.}
    \label{fig:5}
\end{figure}

Ohmic contacts were formed at the four corners of each sample using a
gallium–indium eutectic alloy (Ga:In = 75.5:24.5 by weight, stored under argon atmosphere). The samples were then annealed at $360^{\circ}$C for
$\sim 1$~h to promote alloy infiltration and ensure a stable, low-resistance
interface with the Ge surface. This thermal treatment does not alter the
near-surface properties of HPGe, and previous studies
\cite{raut2020characterization} have shown that such contact formation does not
bias Hall-effect measurements. Net impurity concentrations were measured at
liquid-nitrogen temperature (77 K) in a 0.55 T magnetic field using an Ecopia
HMS-3000 Hall-effect system (Fig.~\ref{fig:5}).

Guided by the known equilibrium segregation behavior of the dominant
impurities \cite{trumbore1960solid}, we interpret the longitudinal profile in
Fig.~\ref{fig:4} as follows. Boron is expected to dominate in the first
$\sim 9$~cm of the crystal, with the combined aluminum and gallium
contribution extending out to $\sim 15$~cm. Phosphorus becomes significant
only in the later portion of the boule, primarily beyond $\sim 15.5$~cm toward
the tail. These regions inform our extraction of effective segregation
coefficients for B, Al+Ga, and P.

\subsection{Methodology}
\label{subsec:method}

The variation of impurity concentration along the crystal during CZ growth can
be described by \cite{pfann1952principles}
\begin{equation}
\label{equation:3.1}
    C = C_0 K_\mathrm{eff} (1-g)^{K_\mathrm{eff}-1},
\end{equation}
where $C$ is the impurity concentration in the solid, $C_0$ is the initial
impurity concentration in the melt, $g$ is the fraction of the melt that has
solidified (obtained from the crystal mass), and $K_\mathrm{eff}$ is the
effective segregation coefficient. Taking the natural logarithm of
Eq.~\eqref{equation:3.1} gives
\begin{equation}
\label{equation:3.2}
    \ln C = (K_\mathrm{eff} - 1)\,\ln(1-g) + \ln(K_\mathrm{eff} C_0\bigr).
\end{equation}

Thus, a plot of $\ln C$ versus $\ln(1-g)$ should yield a straight line whose
slope is $(K_\mathrm{eff} - 1)$ and whose intercept is
$\ln(K_\mathrm{eff} C_0)$. For each dopant region, the experimental data were
fitted using a least-squares linear regression to obtain the slope and
intercept, from which we extract the effective segregation coefficient
$K_\mathrm{eff}$ and the corresponding initial impurity concentration $C_0$.

\section{Results and Discussion}
\label{sec:results_discussion}

The equilibrium distribution coefficient $K_0$ is defined as the ratio of the
impurity concentration in the solid to that in the liquid under equilibrium
conditions, $K_0 = C/C_0$, where $C$ and $C_0$ are the solute concentrations in
the solid and liquid phases, respectively. Traditionally, $K_0$ values are
obtained from phase diagrams via the relative slopes of the solidus and liquidus
lines. Approaching this ideal distribution in practice requires efficient
solute mixing in the melt, controlled by growth rate, crystal rotation and
impurity diffusivity.

\begin{table}[h]
\centering
\caption{Equilibrium segregation coefficients $K_0$ for common impurities in
germanium, defined as the ratio of impurity concentration in the solid to that
in the liquid under equilibrium conditions \cite{trumbore1960solid}.}
\label{tab:seg_coeff}
\begin{tabular}{|c|c|c|c|c|c|}
\hline
\textbf{Impurity} & B & Al & Ga & P & Al--O complex \\
\hline
\textbf{$K_0$} & 17 & 0.073 & 0.087 & 0.080 & $\sim 1$ \\
\hline
\end{tabular}
\end{table}

In real CZ growth, non-equilibrium effects such as melt convection, RF-coil
stirring and axial/radial thermal gradients cause deviations from these
equilibrium values. The relevant quantity for detector-grade crystal pulling is
therefore the effective segregation coefficient $K_\mathrm{eff}$, which
embeds the actual hydrodynamic and thermal conditions. The equilibrium
coefficients $K_0$ for the dominant impurities in Ge, reported by Trumbore
\cite{trumbore1960solid}, are summarized in Table~\ref{tab:seg_coeff}.

Photo-thermal ionization spectroscopy (PTIS) performed at 7 K on the HPGe crystals grown at USD confirms the presence of the primary shallow-level impurities—B, Al, Ga, and P—together with Al–O complexes and hydrogen–oxygen–related defect centers. The stronger PTIS response observed in the dark (blue curve), relative to the illuminated spectrum (red curve) in Figure 5 from \cite{yang2014investigation}, indicates that these centers undergo efficient photo-ionization at cryogenic temperatures, consistent with their behavior as shallow states whose bound carriers are readily depleted under light exposure. The simultaneous observation of donor, acceptor, and complex-related transitions further demonstrates that the material contains a combination of compensating impurities rather than a single dominant species. PTIS is highly sensitive to impurity species, but does not
directly yield absolute concentrations; the relative contributions of the
individual dopants to the net carrier density therefore require complementary
electrical measurements \cite{yang2014investigation}.

\begin{figure}[h]
\centering
    \includegraphics[width=1\textwidth]{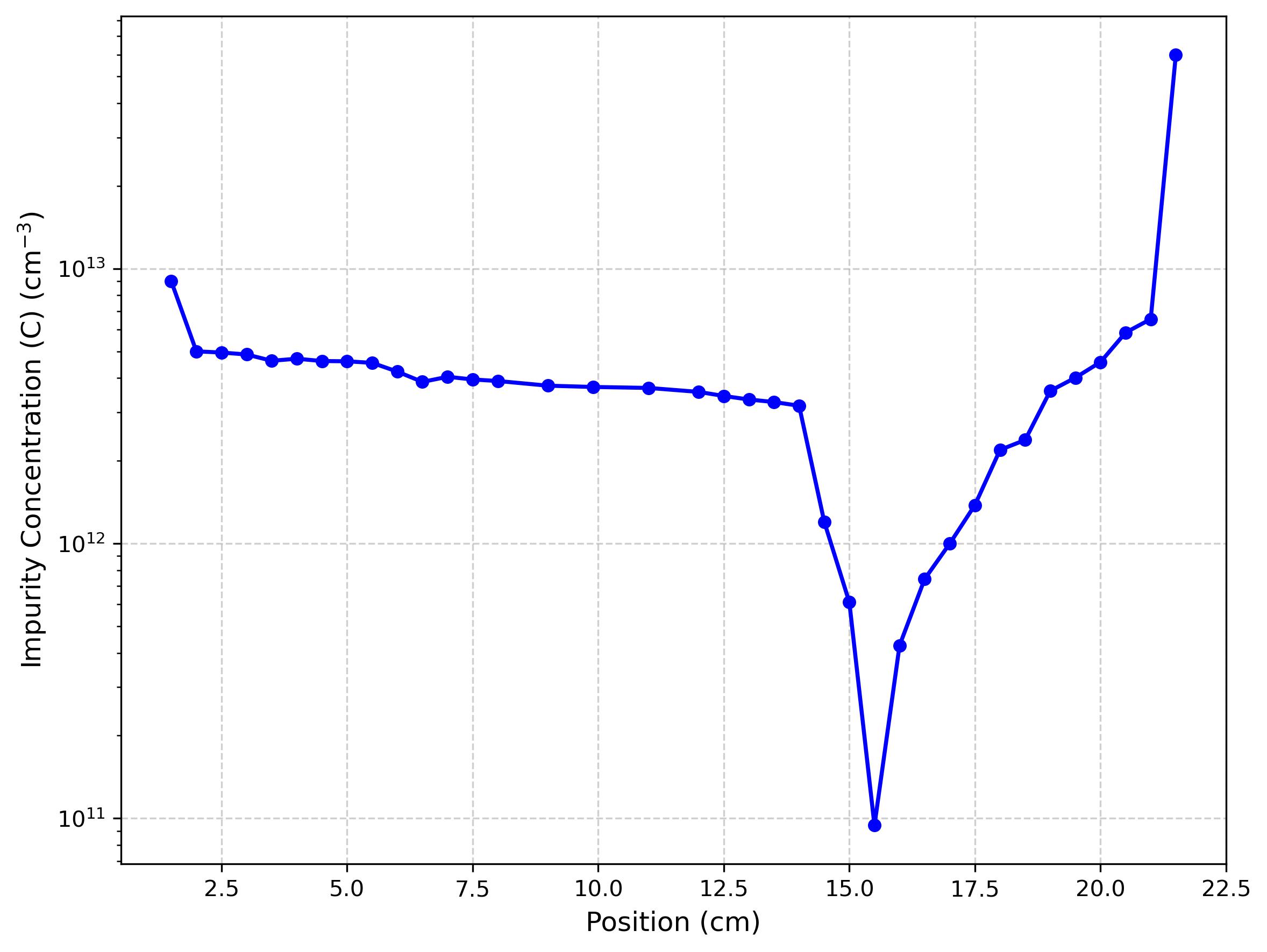}
    \caption{Longitudinal impurity concentration profile of the HPGe crystal
    grown at USD. The crystal is p-type from 1.5 cm to 15 cm and n-type beyond
    15.5 cm. A sharp compensation region between 15.0 and 15.5 cm yields a
    p--n junction.}
    \label{fig:6}
\end{figure}

In this work, we determine the effective segregation coefficients of B, Al+Ga
and P in a CZ-grown HPGe crystal using the impurity-distribution formalism
described in Sec.~\ref{subsec:method}. Bulk impurity concentrations at each
longitudinal position were obtained from Hall-effect measurements at 77 K, and
the resulting profile is shown in Fig.~\ref{fig:6}. The crystal exhibits
p-type behavior from 1.5 cm to 15 cm along its length and becomes n-type
beyond 15.5 cm. A narrow region between 15 cm and 15.5 cm shows sharp
compensation between acceptor (B, Al, Ga) and donor (P) impurities, resulting
in the formation of a p–n junction.

\subsection{Boron and Phosphorus Segregation}

\begin{figure}[h]
\centering
    \includegraphics[width=1\textwidth]{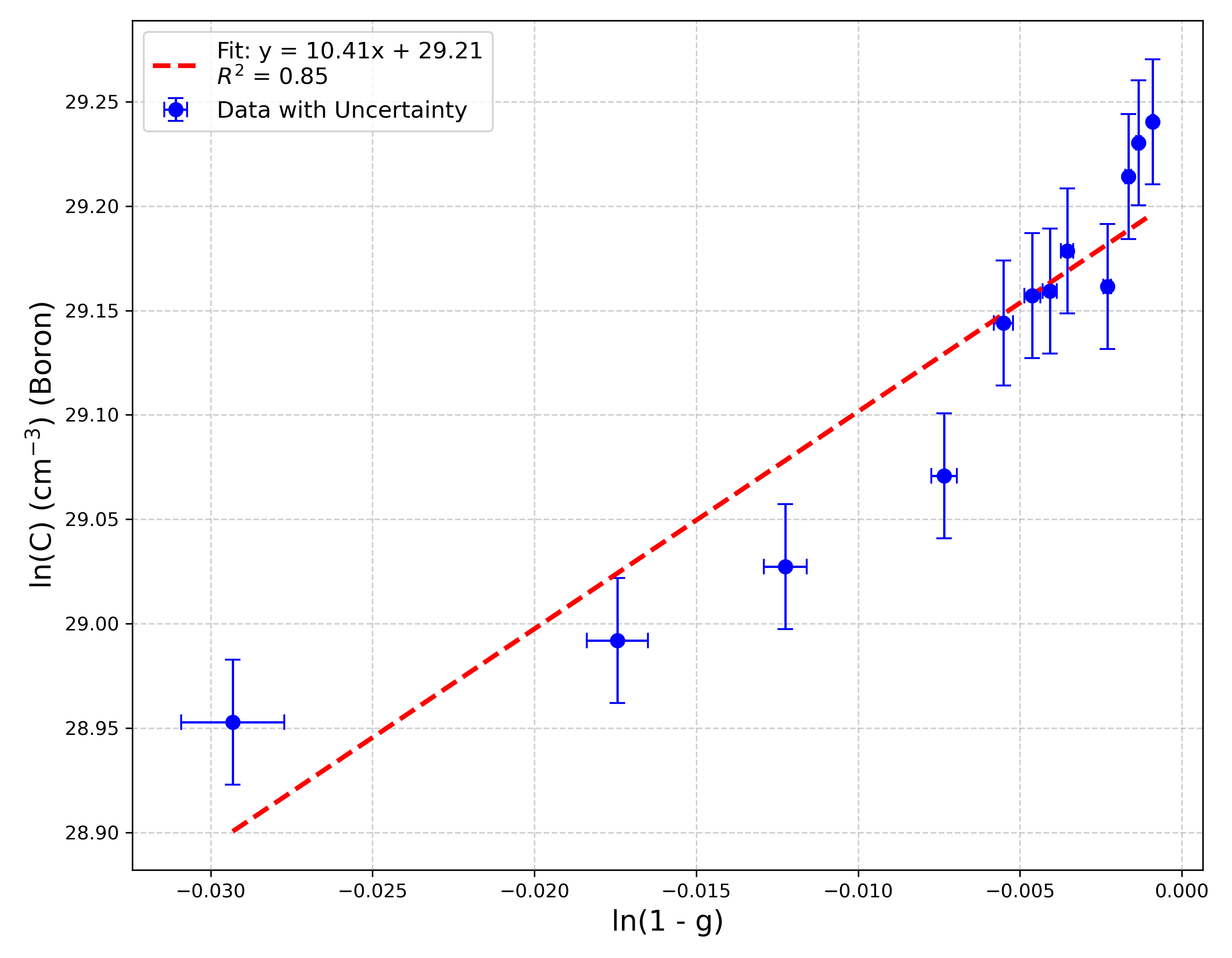}
    \caption{$\ln C$ versus $\ln(1-g)$ for boron, including uncertainties of
    3.04\% in the net carrier concentration and 5.36\% in the solidified
    fraction $g$. Twelve data points from the 2–9 cm region were used. The
    linear regression yields a slope of 10.41 and $R^2 = 0.85$.}
    \label{fig:7}
\end{figure}

\begin{figure}[h]
\centering
    \includegraphics[width=1\textwidth]{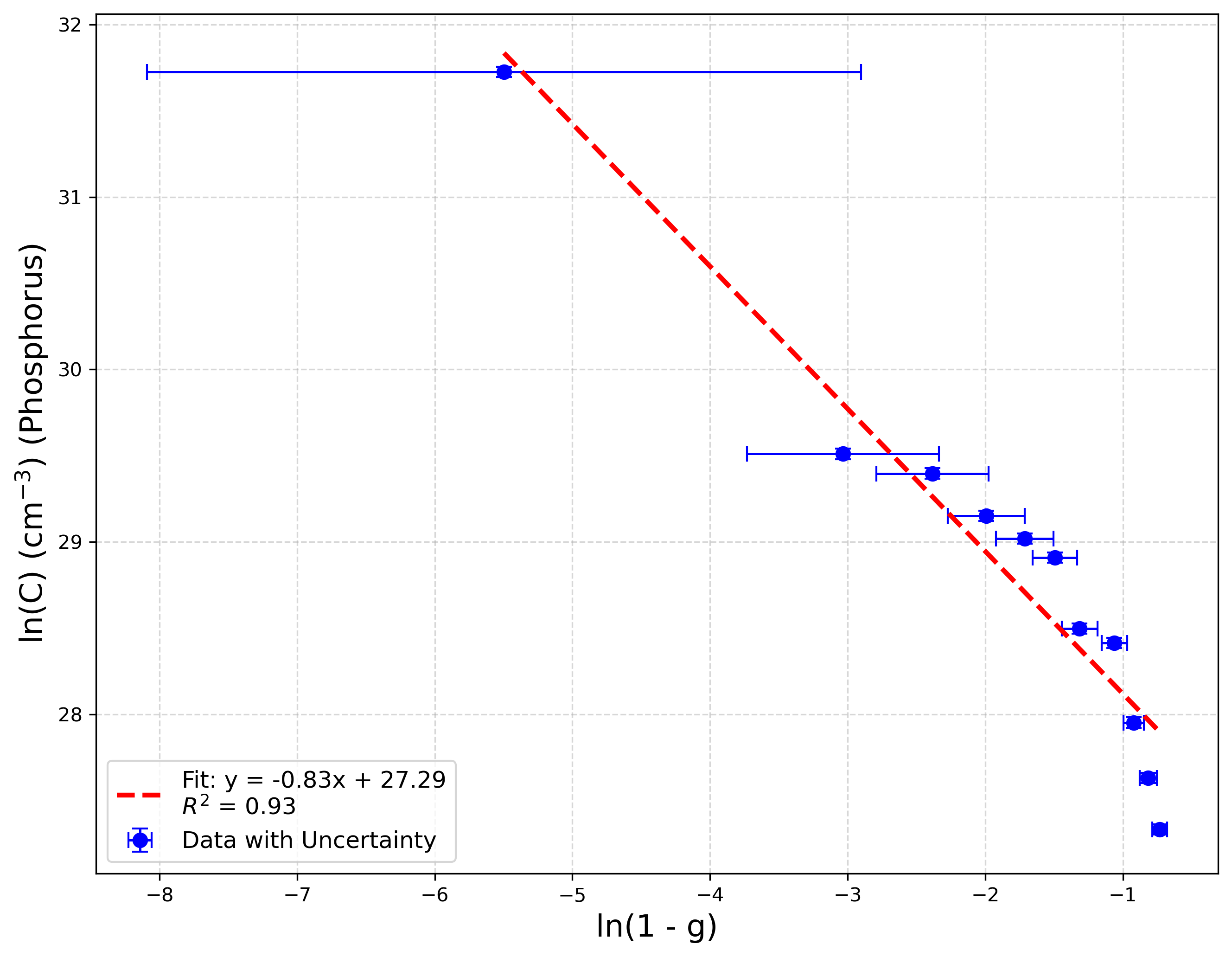}
    \caption{$\ln C$ versus $\ln(1-g)$ for phosphorus, including uncertainties
    of 3.04\% in the net carrier concentration and 5.10\% in $g$. Eleven data
    points from the 16.5–21.5 cm region were used. The linear regression yields
    a slope of $-0.83$ and $R^2 = 0.93$.}
    \label{fig:8}
\end{figure}

The segregation behavior of boron and phosphorus is quantified in the
$\ln C$ versus $\ln(1-g)$ plots shown in Figs.~\ref{fig:7} and \ref{fig:8}.
For B, twelve data points from the 2–9 cm region yield a positive slope in the
$\ln C$--$\ln(1-g)$ plane, indicating that boron is preferentially incorporated
into the solid. For P, eleven data points from the 16.5–21.5 cm region produce
a negative slope, consistent with P remaining predominantly in the melt.

In both cases, a least-squares linear regression was applied to extract the
effective segregation coefficient, where the slope corresponds to
$(K_\mathrm{eff}-1)$ and the intercept to $\ln(K_\mathrm{eff} C_0)$. The B and
P fits yield coefficients of determination $R^2 = 0.85$ and $R^2 = 0.93$,
respectively, demonstrating good agreement with the expected segregation
behavior.

From these fits, the effective segregation coefficients and initial melt
concentrations are
\[
K_\mathrm{eff}^{\mathrm{(B)}} = 11.41 \pm 1.36, \qquad
C_0^{\mathrm{(B)}} = (4.25 \pm 0.06)\times 10^{11}~\mathrm{cm^{-3}},
\]
\[
K_\mathrm{eff}^{\mathrm{(P)}} = 0.17 \pm 0.07, \qquad
C_0^{\mathrm{(P)}} = (3.82 \pm 0.65)\times 10^{12}~\mathrm{cm^{-3}}.
\]

\subsection{Aluminum and Gallium Segregation}

Determining separate $K_\mathrm{eff}$ values for Al and Ga from Hall data alone
is challenging, because the crystal is strongly B-dominated up to about 15 cm,
leading to a pronounced p-type region in which Al and Ga act as secondary
acceptors. The Hall-measured net carrier concentration at position $x_i$,
\begin{equation}
C_{\mathrm{Net}}(x_i) = C_{B}(x_i) + C_{Al}(x_i) + C_{Ga}(x_i) - C_{P}(x_i),
\label{eq:compensation}
\end{equation}
combines the contributions from all acceptors (B, Al, Ga) and the donor P
\cite{bhattarai2024investigating}. To isolate the Al+Ga contribution, we first
use the boron fit to estimate $C_B(x_i)$ and then subtract this from the net
carrier concentration in the 9–15 cm region, where the P contribution is still
small. This procedure yields an effective Al+Ga concentration profile.

\begin{figure}[h]
\centering
    \includegraphics[width=1\textwidth]{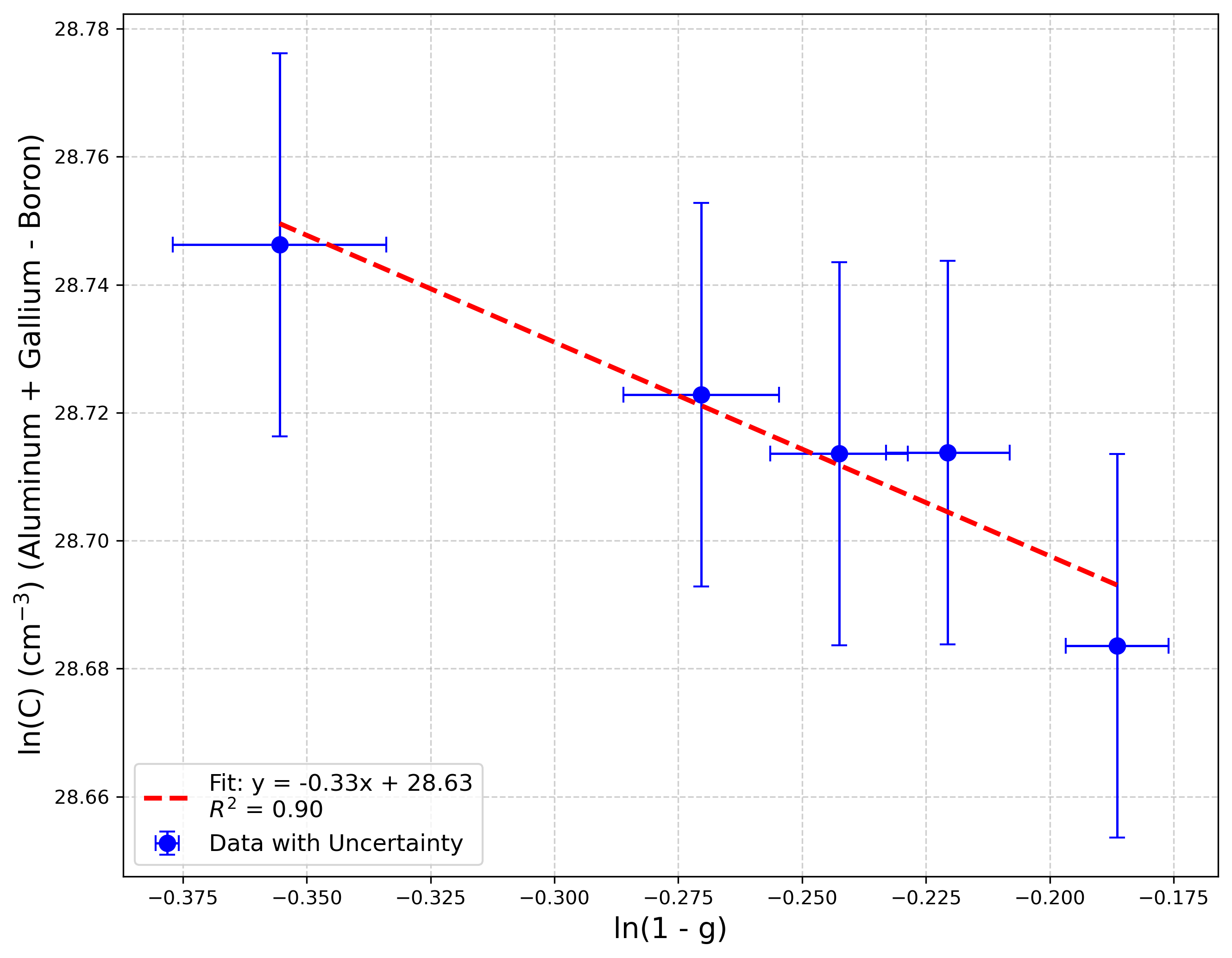}
    \caption{$\ln C$ versus $\ln(1-g)$ for the combined Al+Ga impurity,
    including uncertainties of 3.04\% in the net carrier concentration and
    5.10\% in $g$. The Al+Ga concentration is obtained by subtracting the B
    contribution from the Hall-measured net carrier density. Five data points
    from the 12–14 cm region were used. The fit yields a slope of $-0.33$ and
    $R^2 = 0.90$.}
    \label{fig:9}
\end{figure}

Five data points from the 12–14 cm region were used to construct the
$\ln C$--$\ln(1-g)$ plot for the combined Al+Ga impurity. The resulting linear
fit, shown in Fig.~\ref{fig:9}, yields a negative slope of $-0.33$ with
$R^2 = 0.90$. The corresponding effective segregation coefficient and initial
melt concentration are
\[
K_\mathrm{eff}^{\mathrm{(Al+Ga)}} = 0.67 \pm 0.06, \qquad
C_0^{\mathrm{(Al+Ga)}} = (5.47 \pm 0.09)\times 10^{12}~\mathrm{cm^{-3}}.
\]

The extracted $K_\mathrm{eff}$ and $C_0$ values for all three impurity classes
are summarized in Table~\ref{tab:keff}.

\begin{table}[h]
\centering
\caption{Effective segregation coefficients $K_\mathrm{eff}$ and initial
impurity concentrations $C_{0}$ for B, Al+Ga and P in the HPGe crystal grown at
USD. Uncertainties correspond to 68\% confidence intervals.}
\label{tab:keff}
\begin{tabular}{|c|c|c|}
\hline
\textbf{Impurity} & \textbf{$K_\mathrm{eff}$} & \textbf{$C_{0}$ (cm$^{-3}$)} \\ \hline
Boron & $11.41 \pm 1.36$ & $(4.25 \pm 0.06)\times 10^{11}$ \\ \hline
Aluminum + Gallium & $0.67 \pm 0.06$ & $(5.47 \pm 0.09)\times 10^{12}$ \\ \hline
Phosphorus & $0.17 \pm 0.07$ & $(3.82 \pm 0.65)\times 10^{12}$ \\ \hline
\end{tabular}
\end{table}

\subsection{Uncertainty Analysis}

The Hall-effect measurement setup introduces a systematic uncertainty in the
net carrier concentration of 3.04\% for small, square-shaped samples
\cite{raut2020characterization}. The solidified fraction $g$ is derived from
the mass of each crystal segment and carries additional uncertainty due to
cutting and positioning. A positional deviation of about 1 mm along the boule,
stemming from diamond-wire saw alignment and operator variability, translates
into a thickness and mass uncertainty that varies with crystal diameter.

For B, where the crystal diameter changes most significantly across the measured
regions, the mass differences associated with a 1 mm positional shift were
estimated to be 0.29 g (1.5–2.5 cm), 0.72 g (3–5 cm), 2.02 g (5.5–7.5 cm) and
5.70 g (8–9 cm). These correspond to an average mass-related uncertainty of
5.36\%. For Al+Ga and P, the crystal diameter is more uniform, and the mass
variation over a 1 mm shift is approximately 70 g, giving an average
uncertainty of 5.10\%.

Uncertainties in $K_\mathrm{eff}$ and $C_0$ were obtained via residual analysis
of the linear fits in the $\ln C$--$\ln(1-g)$ plane. Residuals were defined as
the differences between measured and model-predicted $\ln C$ values, and the
scatter was quantified through the standard errors of the slope and intercept.
These errors were then propagated to $K_\mathrm{eff}$ and $C_0$, yielding the
68\% confidence intervals reported in Table~\ref{tab:keff}.

\subsection{Comparison with BPS Model and Literature}

The classic Burton–Prim–Slichter (BPS) model \cite{burton1953distribution}
provides an approximate expression for the effective segregation coefficient:
\begin{equation}
K_{\mathrm{eff}} = 
\frac{K_{0}}{K_{0} + (1 - K_{0}) \exp(-f \delta / D)},
\label{eq:4.2}
\end{equation}
where $K_{0}$ is the equilibrium segregation coefficient, $D$ is the impurity
diffusion coefficient in the melt, $\delta$ is the solute boundary-layer
thickness, and $f$ is the crystal growth rate. The boundary-layer thickness is
given by
\begin{equation}
\delta = 1.6\,D^{1/3}\,\nu^{1/6}\,\omega^{-1/2},
\label{eq:4.3}
\end{equation}
where $\nu = \mu/\rho$ is the kinematic viscosity of molten Ge
($1.35 \times 10^{-3}\,\mathrm{cm^{2}\,s^{-1}}$), and $\omega$ is the crystal
rotation rate. The impurity diffusion coefficients in molten Ge span a range of
values: for B and Ga, $D \approx 3\times10^{-4}$ and
$7.2\times10^{-5}\,\mathrm{cm^{2}\,s^{-1}}$ \cite{taishi2008segregation,
subramanian2023investigation}, while for P and Al,
$D \approx 5\times10^{-5}$ and $6\times10^{-5}\,\mathrm{cm^{2}\,s^{-1}}$
\cite{kodera1963constitutional,becker2017solidification,becker2018free}. For
the combined Al+Ga impurity, we use the averages of their diffusion
coefficients and $K_0$ values.

Using Eq.~\eqref{eq:4.3}, we obtain diffusion boundary-layer thicknesses of
0.063 cm, 0.038 cm and 0.034 cm for B, Al+Ga and P, respectively. Inserting
these into Eq.~\eqref{eq:4.2} with a rotation rate of 3 rpm and the $K_0$ and
$D$ values above yields approximate BPS-based effective segregation coefficients
of $K_\mathrm{eff}^{\mathrm{(BPS)}} \approx 6.12$ for B,
$0.10$ for Al+Ga, and $0.11$ for P.

These estimates are broadly consistent with the trends observed in our data
($K_\mathrm{eff}^{\mathrm{(B)}} \gg 1$, $K_\mathrm{eff}^{\mathrm{(P)}} < 1$),
but differ quantitatively, particularly for Al+Ga, where our measured
$K_\mathrm{eff}^{\mathrm{(Al+Ga)}} = 0.67 \pm 0.06$ is substantially larger
than the BPS estimate. Part of this discrepancy likely arises from the
sensitivity of the BPS model to input parameters that are themselves uncertain,
including $D$, $K_0$ and the exact hydrodynamic conditions in the melt. In
addition, Al- and Ga-related complexes (e.g. Al–O) and incomplete melt
homogenization can alter the effective segregation behavior relative to the
idealized assumptions of the BPS framework.

Earlier experimental work also reports a spread in effective segregation
coefficients. Burton \emph{et al.} \cite{burton1953distribution} obtained
approximate $K_\mathrm{eff}$ values of 0.12 for P and 0.10 for Ga in Ge.
Edwards \cite{edwards1970distribution} later measured $K_\mathrm{eff} \approx
12.2$ for B, emphasizing that slow growth and high rotation rates favor values
closer to $K_0$. Bridgers and co-workers \cite{bridgers1956distribution}
reported a wider range for B (2.2–17.4), with higher values correlated with
increased growth rates, underscoring the impact of non-equilibrium conditions.

Within this broader context, the present results show internally consistent
$K_\mathrm{eff}$ values across all three impurity classes for a single crystal
grown under well-documented conditions. They thus provide a more tightly
constrained benchmark for impurity segregation in detector-grade HPGe grown by
the Czochralski method and offer practical input for tuning growth parameters
to achieve desired longitudinal impurity profiles.

\section{Conclusion}

High-purity germanium (HPGe) crystals were successfully grown at USD using the
Czochralski technique, with diameters ranging from $\sim 8$ cm near the
shoulder to $\sim 11$ cm toward the tail. For the first time at USD, a
detector-grade HPGe crystal was longitudinally sectioned into 37 segments,
enabling high-resolution mapping of net impurity concentrations via Hall-effect
measurements at 77 K. Using a standard impurity-distribution model and
ln–ln linear regression, we extracted effective segregation coefficients for
boron, combined aluminum and gallium, and phosphorus of
$11.41 \pm 1.36$, $0.67 \pm 0.06$, and $0.17 \pm 0.07$, respectively. The
corresponding initial melt concentrations were
$(4.25 \pm 0.06)\times 10^{11}$, $(5.47 \pm 0.09)\times 10^{12}$, and
$(3.82 \pm 0.65)\times 10^{12}~\mathrm{cm^{-3}}$.

These values provide a quantitative, internally consistent picture of impurity
segregation under realistic HPGe detector-growth conditions at USD and
complement classical Burton–Prim–Slichter estimates and prior literature.
The methodology demonstrated here establishes a reproducible framework for
linking growth parameters to longitudinal impurity profiles and will guide
optimization of zone refining and CZ pulling to increase the yield and
uniformity of detector-grade HPGe. In turn, this supports the development of
next-generation low-background HPGe detectors for rare-event physics
applications, including dark-matter and neutrinoless double-beta decay
searches.

\acknowledgments
This work was supported in part by NSF OISE 1743790, NSF PHYS 2117774, NSF OIA 2427805, NSF PHYS 2310027, NSF OIA 2437416, DOE DE-SC0024519, DE-SC0004768, and a research center supported by the State of South Dakota. \cite{mei2025probing}

\section*{Author Contributions}
The conceptualization and methodology were developed by S. Chhetri, D.-M. Mei and S. Bhattarai. Crystal growth was performed by S. Chhetri, S. Bhattarai, N. Budhathoki, A. Warren, K.-M. Dong. Sample preparation was supported by  S. A. Panamaldeniya, and A. Prem. Experimental measurements were carried out by S. Chhetri. Data curation, formal analysis, and the original draft were completed by S. Chhetri. Visualization and manuscript review were carried out by S. Chhetri and D.-M. Mei. Project supervision, resource provision, and funding acquisition were led by D.-M. Mei. All authors contributed to the development of the study and approved the final manuscript.


\bibliographystyle{JHEP}
\bibliography{biblio.bib}






\end{document}